\documentclass[unsortedaddress,twocolumn,showpacs,preprintnumbers,amsmath,amssymb]{revtex4}
\usepackage{graphicx}% Include Figure files
\usepackage{dcolumn}% Align table columns on decimal point
\usepackage{bm}% bold math
\usepackage{latexsym}
\begin{document}

\vspace{30mm}

\title{Analysis of music: controlled random music and probability distribution function of recurrence time of amplitude peaks}

\author{Vishnu Sreekumar{\footnote{Electronic mail: Vishnu.Sreekumar85@gmail.com}}}
\author{Mahendra K Verma}
\affiliation{Department Of Physics, Indian Institute Of Technology, Kanpur 208016, Uttar Pradesh, India}
%{\footnote{Corresponding author; e-mail: mkv@iitk.ac.in}}
%\author{Mahendra K Verma}
\author{Venkatesh K S}

\affiliation{Department Of Electrical Engineering, Indian Institute Of Technology, Kanpur 208016, Uttar Pradesh, India}

%\affiliation{Department Of Physics, Indian Institute Of Technology, Kanpur 208016, India}

%\affiliation{Physique Statistique et Plasmas, CP231, Facult\'e des Sciences, Universit\'e Libre de Bruxelles, B-1050
%Bruxelles, Belgium}

\author{Gaurav Narain}

\affiliation{SISSA, via Beirut 4, 34014 Trieste, Italy}

\date {23 November, 2008}% 

%%%%%%%%%%%%%%%%%%%%%%%%%%
%%%%%%%%%%%%%%%%%%%%%%%%%%
\begin{abstract}
Correlations in music that exist within its waveform are studied. Monophonic wave files of random music are generated and the probability distribution function of time interval between large signal values is analyzed. A power law behavior for the distribution function in the range from 0.1millisecond to 20 milliseconds is observed. An attempt is made to investigate the origin of these correlations by randomizing each of the factors (frequencies, intensities and durations of notes of the random music files) separately.

\end{abstract}

%%%%%%%%%%%%%%%%%%%%%%%%%%

%\pacs{91.25.Cw, 47.65.Md, 05.45.Ac}
\maketitle
%%%%%%%%%%%%%%%%%%%%%%%%%%

%%%%%%%%%%%%%%%%%%%%%%%%%%%%%%%%%%%%%%%%%%%%%%%%%%%%%%%%%%%%%%%%%
\section*{\label{sec1} I. INTRODUCTION}
%%%%%%%%%%%%%%%%%%%%%%%%%%%%%%%%%%%%%%%%%%%%%%%%%%%%%%%%%%%%%%%%
 Trying to quantify any form of art is a challenging task. The attempt here is to look into the waveform of music using statistical and computational methods and try to investigate the correlations that exist in music. Various studies have been done prior to this which analyze correlations in music. Voss and Clarke1 did pioneering work in this field. Voss and Clarke analyzed the spectrum of time series of various genre of music including Classical, Jazz, and Rock taken from radio broadcasts. The durations of the samples extended to about 12 hours each. They observed prominent $1/f$ spectral behavior in all the time series that they analyzed. They came to a conclusion that all ``intelligent'' behavior should show a $1/f$ like spectral density.
 
Beran \cite{ref2} argues that $1/f$ behavior could be due to the instruments alone and not depend on the composers or the genre. He demonstrates that even one single note shows a $1/f$ noise-like behavior. A whole musical piece is considered by most musicians to be the largest unit of artistic significance \cite{ref3,ref4}. Since Voss and Clarke had included hours of recording from radio stations that included announcements and played different pieces of music, the analysis done may not be conclusive enough to make deductions about music \cite{ref5} and how to quantify its aesthetic attributes. Boon and Decroly \cite{ref3} analyzed 23 pieces and got the value of the slope of the bi-logarithmic plot of the power spectrum to be consistently lying in between 1.79 and 1.97 which is close to red noise-like behavior. Their results also matched Nettheim's results \cite{ref4}. Both groups analyzed pieces of music which are considered to be the largest units of significance of musical expression. Other attempts to computationally identify aspects of musicality have been made in the past. Manaris {\it et al.} \cite{ref6} found that certain musical attributes exhibit the Zipf-Mandelbrot distribution (a power law distribution on ranked data) and that statistical analysis of certain metrics defined on music can potentially be reliable in computationally identifying aesthetic attributes of music. 

Voss and Clarke also applied their results to stochastic composition. This was however done using two separate melody and duration sequences which were both assumed to be separate $1/f$ processes. The meaning of a spectrum of durations is unclear, for a duration itself moves along the time axis \cite{ref4}. Rhythm has been converted into more meaningful data sets by Su and Wu \cite{ref7}, and studies on clustering patterns of these data points have been done to understand the fractal nature of music \cite{ref8,ref9}. Fractal dimension plots have also shown agreement with perception density of events during a piece of music \cite{ref10}. The multifractal spectrum plot also seems to be a potential tool for computationally distinguishing various styles of music \cite{ref7}. The objective in this paper is to look for structure in terms of correlations in peaks within the waveform and make an attempt to look for other statistical features than the ones described above that may be intrinsic to music.
 
 It has been found that in many natural non-equilibrium systems, the probability distribution of recurrence time of the peaks (or large events) follow power laws hence showing self-similarity properties \cite{ref11,ref12}. 
 
\begin{equation}
P(\tau ) = A \tau^{-\beta}  
\end{equation}
where $\tau$ is the interval between large events, and P($\tau$) is the interval distribution. The above law is due to the temporal correlations in the system. These features have been observed for earthquakes \cite{ref13}, solar flares \cite{ref14}, and many other phenomena. Verma {\it et al.} \cite{ref15} derived a universal scaling law connecting the sizes, recurrence time, and spatial interval of large events and applied it to voltage-dependent anion channels (VDAC) current time series. Inspired by such studies that analyze correlations in recurrence time of large events, in the present manuscript the same idea has been applied to music.  P. Diodati and S. Piazza \cite{ref16} had done a similar study with a focus on Gioacchino Rossini's La calunnia `e un venticello from Il barbiere di Siviglia. We probe the origin of the correlations by systematically varying the musical parameters.

The organization of the paper is as follows: In Section II several time series of computer-generated music are analyzed. The results are described for the statistics of the duration between large peaks.  Section III contains conclusions.

%%%%%%%%%%%%%%%%%%%%%%%%%%%%%%%%%%%%%%%%%%%%%%%%%%%%%%%%%%%%%%%%%
\section*{\label{sec2} II. ANALYSIS OF CONTROLLED RANDOM MUSIC}
%d%%%%%%%%%%%%%%%%%%%%%%%%%%%%%%%%%%%%%%%%%%%%%%%%%%%%%%%%%%%%%%%%

Music time series are generated using a computer program and analyzed.  The output of the program is in MIDI format, which in turn is converted into WAVE format for analysis. Short music files of 50 notes each are first generated and given random characteristics of rhythm (durations of notes), melody (frequencies of the notes) and loudness (intensities with which the notes were played). This (50 notes) is probably too short to be considered as a significant piece of musical expression (though monophonic mobile phone ring tones could be only just almost as long). Music files are then generated with the same combinations but now with 300 notes and with a different (panflute) tone. This is closer to a unit of meaningful musical expression. Only monophonic music is studied here.

 These time series are created with varying degrees of randomness of intensities, frequencies and durations of notes. For example, one file has notes played with the same intensities, and same durations, but different frequencies. All other such combinations are studied. Probability distributions are assigned from which the intensities and frequencies are selected.  Durations are randomly assigned but have to be chosen from a restricted set of durations defined corresponding to a 32nd note (demi semi quaver) as $2^{0}$ times 100ms, 16th note (semi quaver) as $2^{1}$ times 100ms, 8th note (quaver) as $2^{2}$ times 100ms, quarter note (crotchet) as $2^{3}$ times 100ms, half note (minim) as $2^{4}$ times 100ms and a whole note (semi breve) as $2^{5}$ times 100ms.

We analyze computer-generated violin and panflute music time series by varying three parameters, frequencies, intensities and durations. These three parameters could either be random or held constant. The random frequencies are chosen from two different probability distributions, quadratic random or uniform random. The quadratic distribution is given by
\begin{equation}
P(f) = B(f - f_{min} )(f_{max} - f)  
\end{equation}
where B=$10^{-4}$ , f is the frequency, $f_{min}$ and $f_{max}$ are the minimum and maximum values of f. They are denoted by the numbers 36 and 96 respectively in the computer program. The number 60 denotes the `middle C'' which is about 261 Hz. This quadratic probability distribution peaks at the average value of $f_{min}$ and $f_{max}$. Frequencies from the middle of this defined range of values are more often selected when the quadratic distribution is used. The uniform random distribution is defined for the same range of frequencies. All frequencies in the defined range are equally probable to be chosen when the uniform distribution is used. In this paper, these two distributions are denoted by Qr and Ur respectively. The intensities are taken to be either constant, or random with quadratic distribution. Similarly durations are either constant or randomly chosen from the set defined in the previous paragraph. For brevity we denote the time series by a short phrase whose first two letters denote the nature of randomness of the frequencies, the third and the fourth letters denote the same for the intensities and durations respectively. Here S denotes ``same'' and R denotes ``random''. See Table 1 and Table 2 for the representation of violin and panflute music time series. For example, a time series named QrRS contains notes of random frequencies (chosen from a quadratic distribution), random intensities and same frequencies.

  %%%%%%%%%%%%%%%%%%%%%%%%%%%%%%%%%%%%%%%%%%%%%%%%%%%%%%%%%%%%%%%%%
\subsection* {A. Violin plots}
%%%%%%%%%%%%%%%%%%%%%%%%%%%%%%%%%%%%%%%%%%%%%%%%%%%%%%%%%%%%%%%%%
Here we analyze the time series of the computer generated violin music. The wave file is read and 1006970 data points are used for this analysis. The music files (violin) are 22 seconds long and contain 50 notes each. Hence the average duration of one note is about 440 ms.  For the time series, a threshold of 1.5$\sigma$ is chosen as a cut off for identifying the peaks or the large events. Here $\sigma$ is the standard deviation. After this, the interval distribution is computed. Fig. 1-4 illustrate the distributions for four different cases of Table 1. In Fig. 1-4, unit time interval is 22/1006970 = 21.8 $\mu$seconds.

The plots exhibit the power law behavior for the recurrence interval between the peaks. The slopes range from -1.6 to -2.7. Note that P($\tau$) $\sim$ $\tau^{-2}$ implies 1/f spectral density when the amplitude of the peaks are uniform \cite{ref17}. Since the slopes described here are close to -2, these results are in general agreement with 1/f spectral densities reported by Voss and Clarke. For completeness, in Tables 1 and 2 we also report $R^{2}$ that denotes the proportion of the variation that is explained by the model, which is the power law model in this case. 
%%%%%%%%%%%%%%%%%%%%%%%%%%%%%%%%%%%%%%%%%%%%%%%%%%%%%%%%%%%%%%%%%%%%%% 
\begin{table}[h!]
\caption{Slopes and coefficients of determination ($R^{2}$) for the bi-logarithmic plots of the probability distribution function of time interval between large signal values for the violin wavefiles.}
\begin{center}
\begin{tabular}{|c|c|c|}
\hline
\textbf{Violin Time Series}   &   \textbf{Slope}  & \textbf{$R^{2}$}  \\
\hline
Fig.1 QrSS & -2.0 & 0.7362 \\
Fig.2 UrSS & -1.6 & 0.6905 \\
Fig.3 UrRS & -2.5 & 0.9605 \\
Fig.4 UrSR & -2.7 & 0.9605 \\
\hline
\end{tabular}
\end{center}
\end{table}
%%%%%%%%%%%%%%%%%%%%%%%%%%%%%%%%%%%%%%%%%%%%%%%%%%%%%%%%%%%%%%%%%%%%%%%
                                                                                
%%%%%%%%%%%%%%%%%%%%%%%%%%%%%%%%%%%%%%%%%%%%%%%%%%%%%%%%%%%%%%
\begin{figure}[h!]
\begin{center}
\includegraphics[width=0.78\columnwidth]{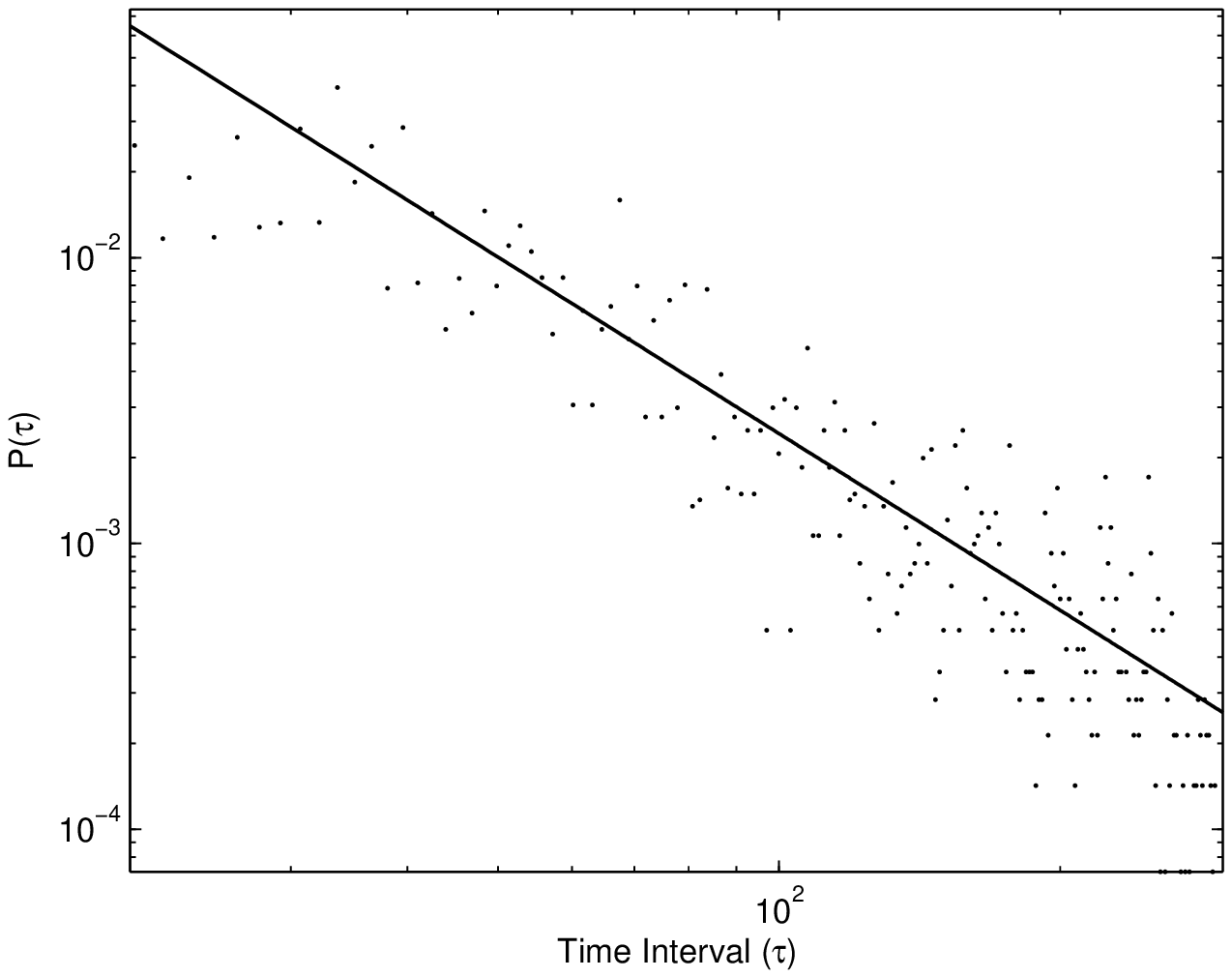}
\end{center}
\caption{QrSS
}
\label{fig1}
\end{figure}
%%%%%%%%%%%%%%%%%%%%%%%%%%%%%%%%%%%%%%%%%%%%%%%%%%%%%%%%%%%%%
%%%%%%%%%%%%%%%%%%%%%%%%%%%%%%%%%%%%%%%%%%%%%%%%%%%%%%%%%%%%%
\begin{figure}[h!]
\begin{center}
\includegraphics[width=0.78\columnwidth]{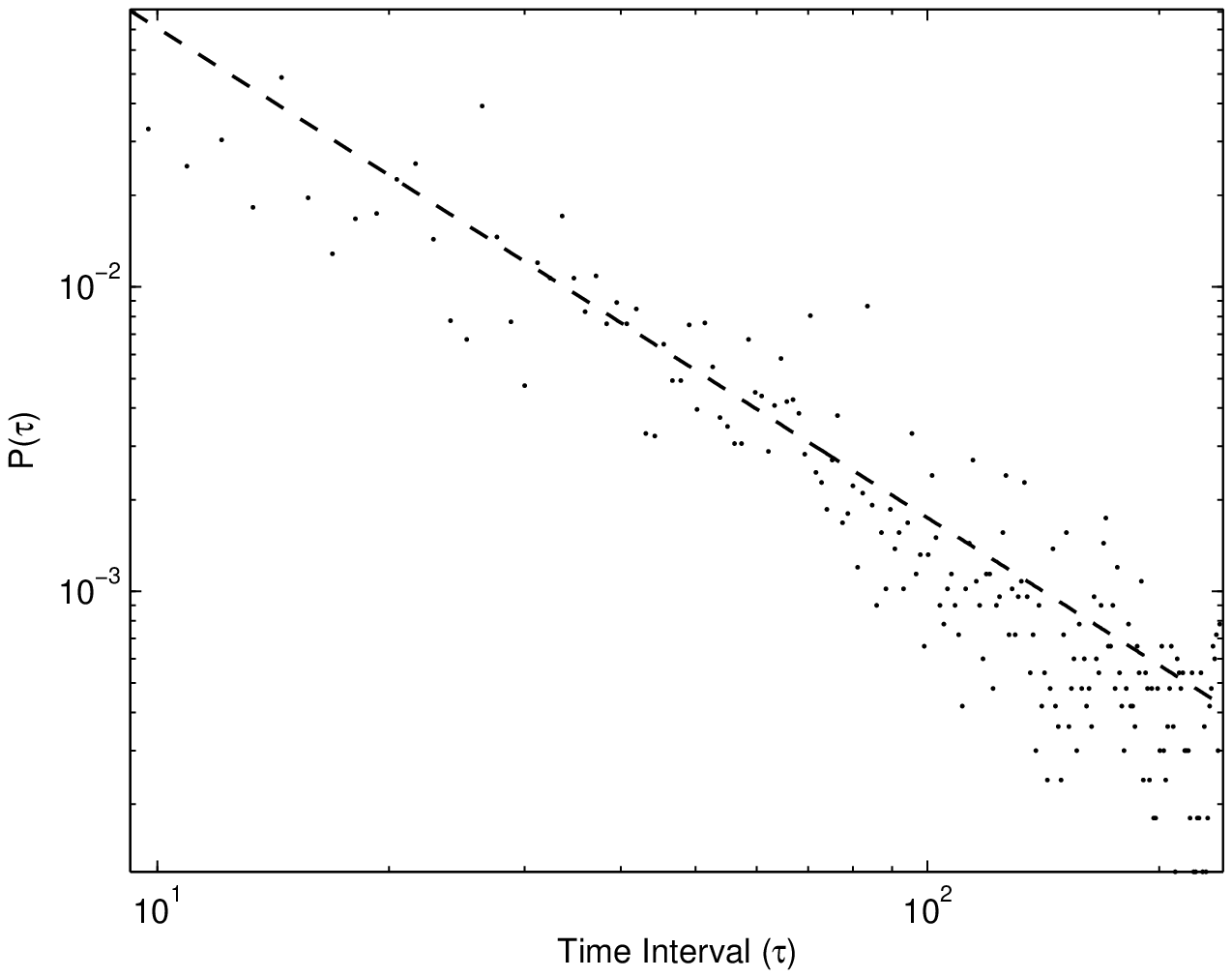}
\end{center}
\caption{UrSS
}
\label{2}
\end{figure}
%%%%%%%%%%%%%%%%%%%%%%%%%%%%%%%%%%%%%%%%%%%%%%%%%%%%%%%%%%%%%
%%%%%%%%%%%%%%%%%%%%%%%%%%%%%%%%%%%%%%%%%%%%%%%%%%%%%%%%%%%%%
\begin{figure}[h!]
\begin{center}
\includegraphics[width=0.78\columnwidth]{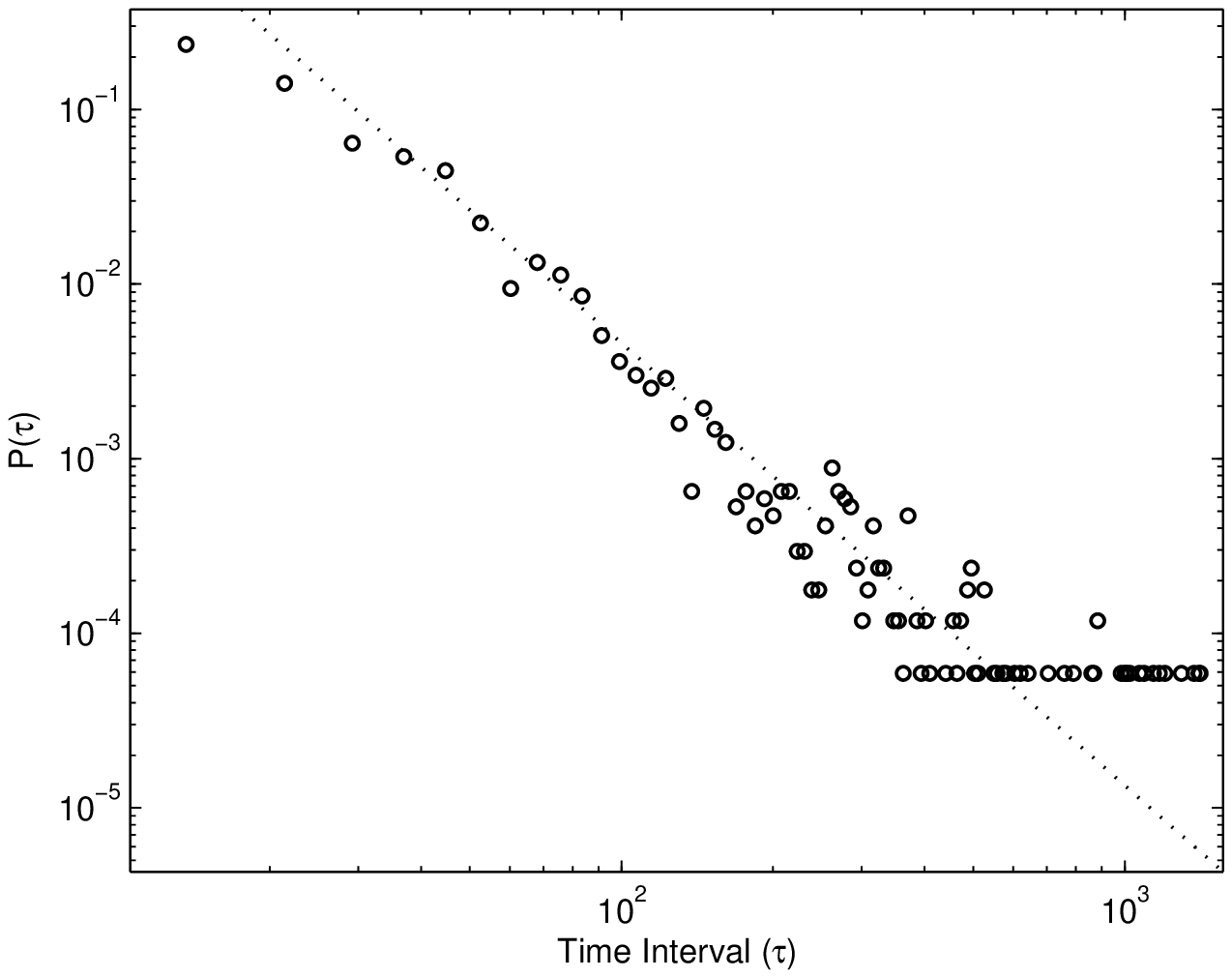}
\end{center}
\caption{UrRS
}
\label{3}
\end{figure}
%%%%%%%%%%%%%%%%%%%%%%%%%%%%%%%%%%%%%%%%%%%%%%%%%%%%%%%%%%%%%
%%%%%%%%%%%%%%%%%%%%%%%%%%%%%%%%%%%%%%%%%%%%%%%%%%%%%%%%%%%%%
\begin{figure}[h!]
\begin{center}
\includegraphics[width=0.78\columnwidth]{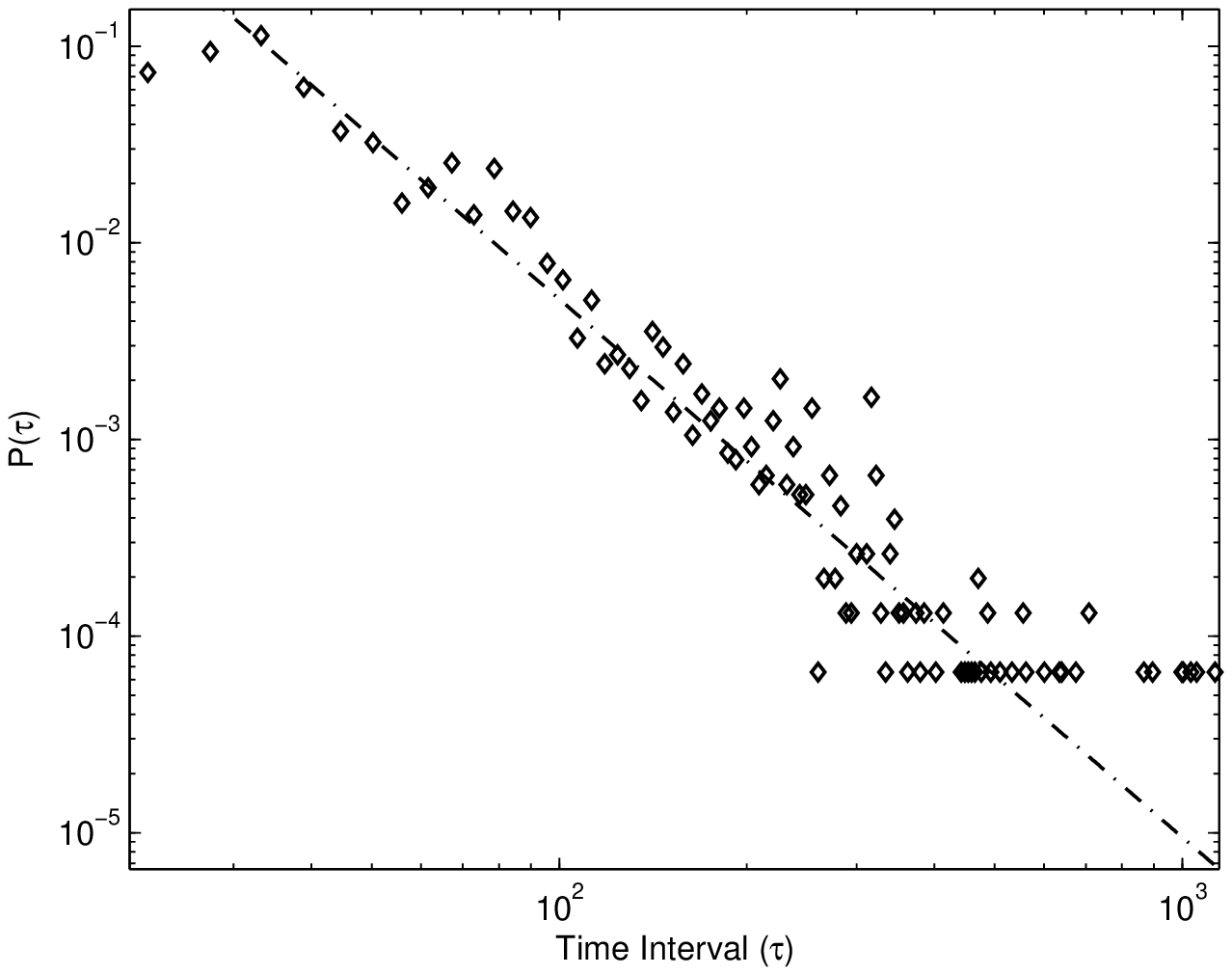}
\end{center}
\caption{UrSR
}
\label{4}
\end{figure}
%%%%%%%%%%%%%%%%%%%%%%%%%%%%%%%%%%%%%%%%%%%%%%%%%%%%%%%%%%%%%

%%%%%%%%%%%%%%%%%%%%%%%%%%%%%%%%%%%%%%%%%%%%%%%%%%%%%%%%%%%%%%%%%
\subsection* {B. Panflute plots}
%%%%%%%%%%%%%%%%%%%%%%%%%%%%%%%%%%%%%%%%%%%%%%%%%%%%%%%%%%%%%%%%%
 Panflute music time series are studied here. 6978845 data points are used for the analysis. The computer generated panflute music files are 156 seconds long and contain 300 notes each. Hence the average duration of one note is about half a second. As in the case for the violin time series, a threshold of 1.5$\sigma$ is chosen as cutoff and the peaks are identified. Fig. 5-7 illustrate the interval distributions for three cases of Table 2. Value corresponding to unit time interval in the Fig 5-7 is 156/6978845 = 22.35 $\mu$seconds. The interval distributions exhibit the power law behavior and the slopes range from -1.5 to -2.2. 
%%%%%%%%%%%%%%%%%%%%%%%%%%%%%%%%%%%%%%%%%%%%%%%%%%%%%%%%%%%%%%%%%
\begin{table}[h!]
\caption{Slopes and coefficients of determination ($R^{2}$) for the bi-logarithmic plots of the probability distribution function of time interval between large signal values for the panflute wavefiles.}
\begin{center}
\begin{tabular}{|c|c|c|}
\hline
\textbf{Panflute Time Series}   &   \textbf{Slope}  & \textbf{$R^{2}$}  \\
\hline
Fig.5 QrSS & -1.5 & 0.6501 \\
Fig.6 UrSS & -1.9 & 0.7642 \\
Fig.7 UrRS & -2.2 & 0.9970 \\
\hline
\end{tabular}
\end{center}
\end{table}
%%%%%%%%%%%%%%%%%%%%%%%%%%%%%%%%%%%%%%%%%%%%%%%%%%%%%%%%%%%%%%%%%                                                                                
%%%%%%%%%%%%%%%%%%%%%%%%%%%%%%%%%%%%%%%%%%%%%%%%%%%%%%%%%%%%%%
\begin{figure}[h!]
\begin{center}
\includegraphics[width=0.78\columnwidth]{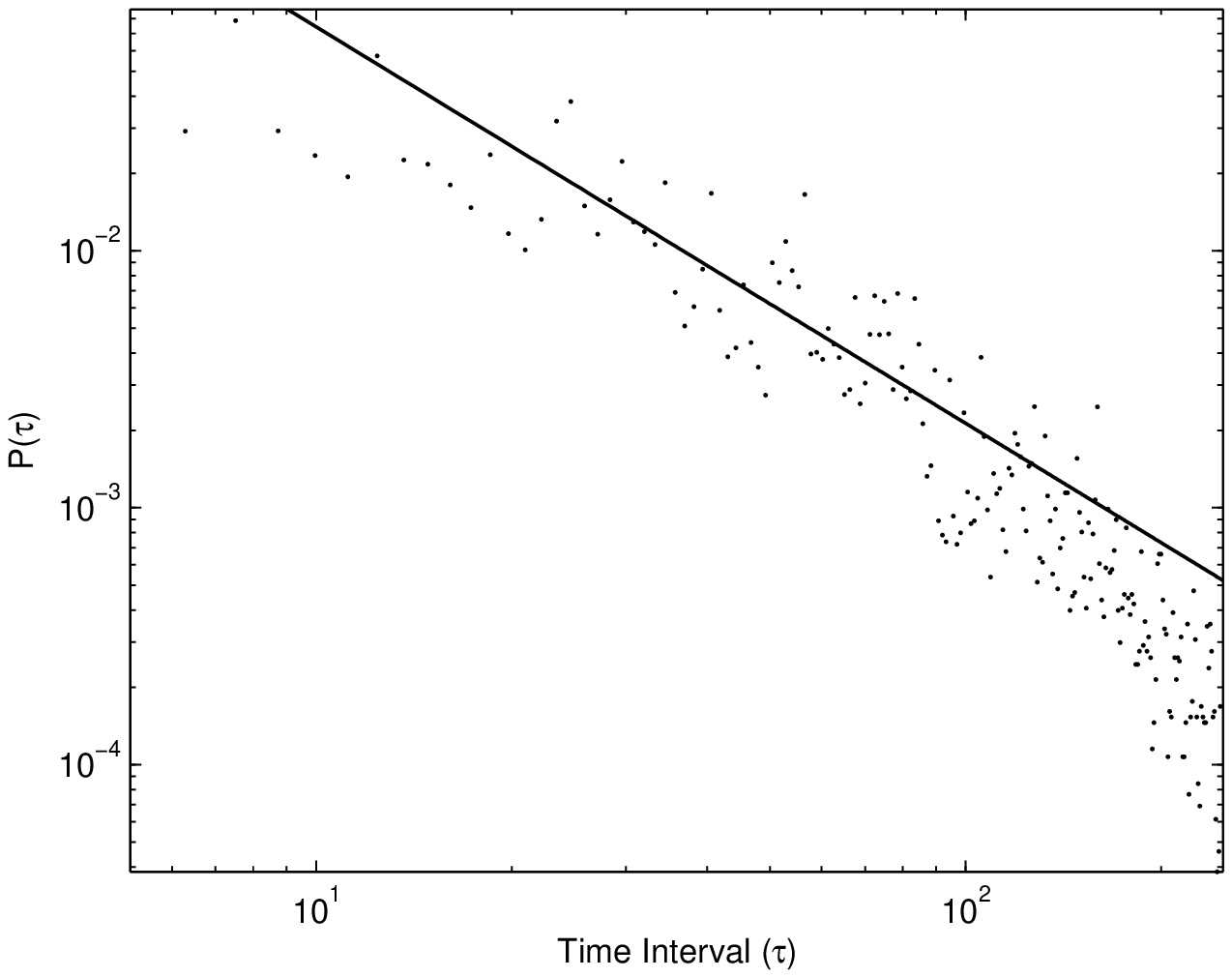}
\end{center}
\caption{QrSS}
\label{5}
\end{figure}
%%%%%%%%%%%%%%%%%%%%%%%%%%%%%%%%%%%%%%%%%%%%%%%%%%%%%%%%%%%%%
%%%%%%%%%%%%%%%%%%%%%%%%%%%%%%%%%%%%%%%%%%%%%%%%%%%%%%%%%%%%%
\begin{figure}[h!]
\begin{center}
\includegraphics[width=0.75\columnwidth]{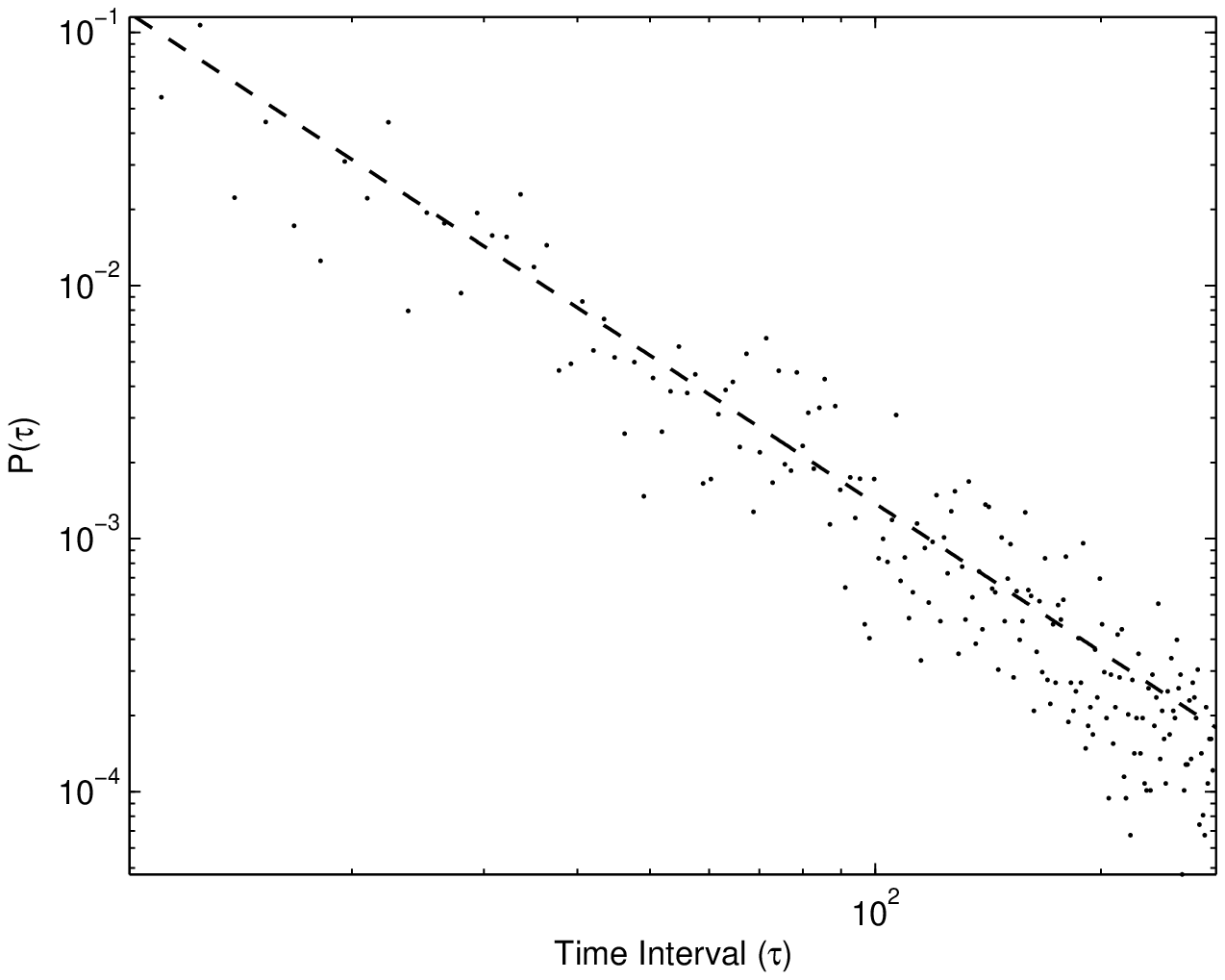}
\end{center}
\caption{UrSS}
\label{6}
\end{figure}
%%%%%%%%%%%%%%%%%%%%%%%%%%%%%%%%%%%%%%%%%%%%%%%%%%%%%%%%%%%%%
%%%%%%%%%%%%%%%%%%%%%%%%%%%%%%%%%%%%%%%%%%%%%%%%%%%%%%%%%%%%%
\begin{figure}[h!]
\begin{center}
\includegraphics[width=0.75\columnwidth]{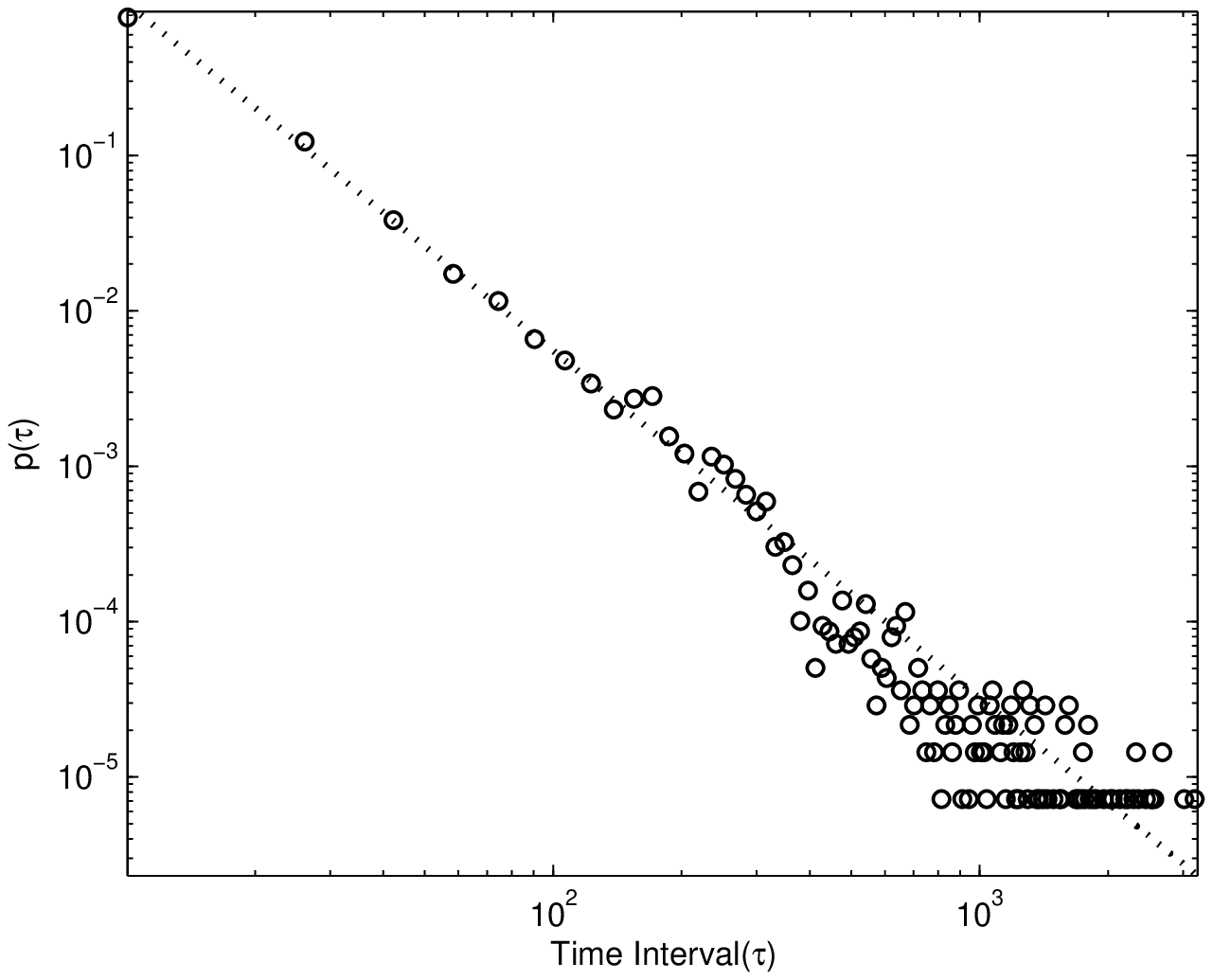}
\end{center}
\caption{UrRS}
\label{7}
\end{figure}
%%%%%%%%%%%%%%%%%%%%%%%%%%%%%%%%%%%%%%%%%%%%%%%%%%%%%%%%%%%%%
 
In the next subsection, we probe the role of the various musical attributes in giving rise to the power law correlations.

%\vspace{5mm}

 %%%%%%%%%%%%%%%%%%%%%%%%%%%%%%%%%%%%%%%%%%%%%%%%%%%%%%%%%%%%%%%%%
\subsection* {C. Repeated notes}
%%%%%%%%%%%%%%%%%%%%%%%%%%%%%%%%%%%%%%%%%%%%%%%%%%%%%%%%%%%%%%%%%
To understand the origin of the power law correlations, the three factors, melody, rhythm and loudness are randomized separately. Frequencies of the notes are held constant (261 Hz) throughout this study while intensities and durations are randomized systematically. The power law disappears completely when all three parameters are held constant, i.e, when the notes are repeated 50 times with the same intensities and durations (Fig. 8). Randomizing durations alone (Fig. 9) does not help restore the power law behavior but the power law reappears when randomization of durations is coupled with randomization of intensities (Fig. 11). Randomizing intensities alone, while holding both frequencies and durations constant also restores the power law behavior (Fig. 10).

%%%%%%%%%%%%%%%%%%%%%%%%%%%%%%%%%%%%%%%%%%%%%%%%%%%%%%%%%%%%%
\begin{figure}[h!]
\begin{center}
\includegraphics[width=0.78\columnwidth]{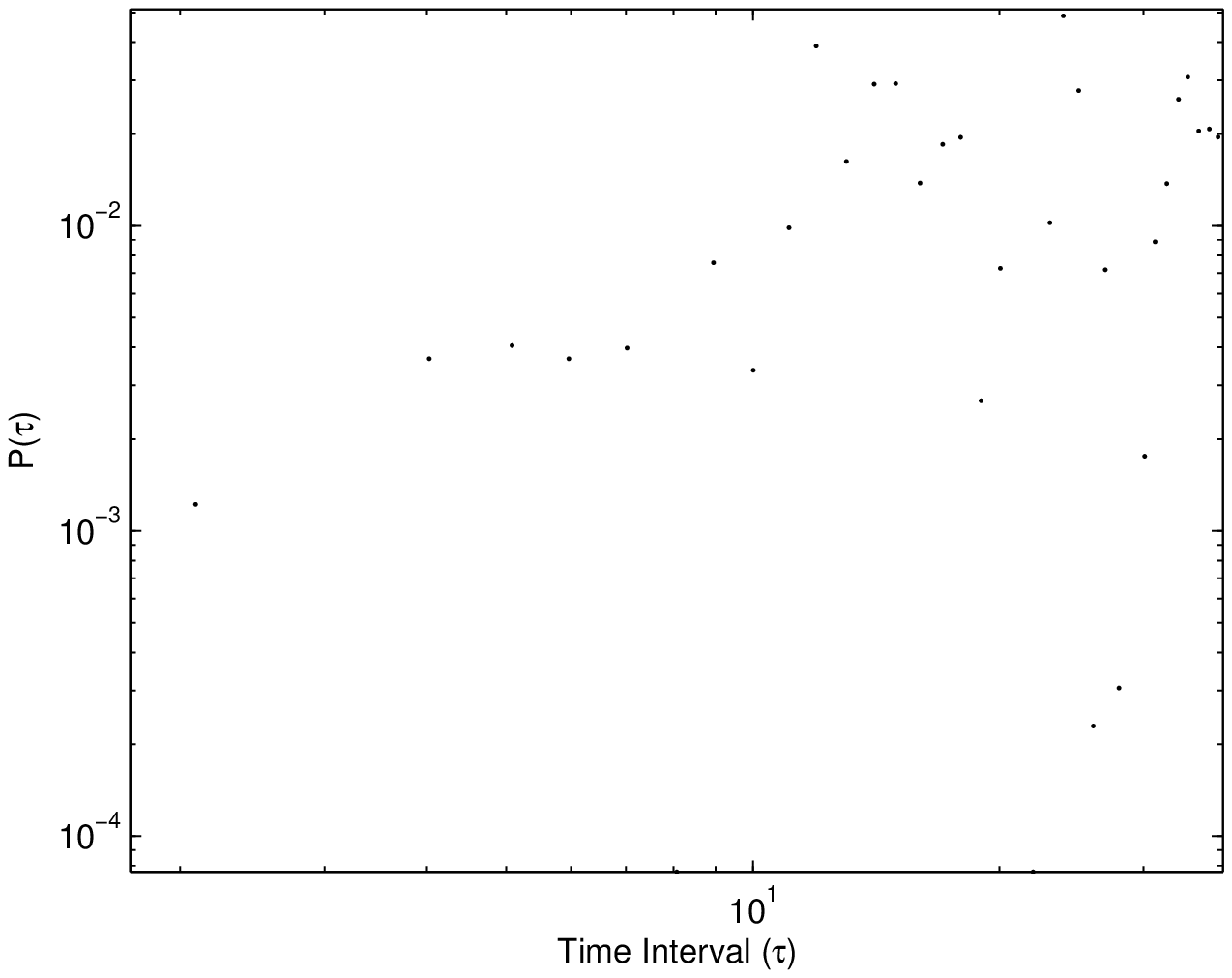}
\end{center}
\caption{SSS}
\label{8}
\end{figure}
%%%%%%%%%%%%%%%%%%%%%%%%%%%%%%%%%%%%%%%%%%%%%%%%%%%%%%%%%%%%%
%%%%%%%%%%%%%%%%%%%%%%%%%%%%%%%%%%%%%%%%%%%%%%%%%%%%%%%%%%%%%
\begin{figure}[h!]
\begin{center}
\includegraphics[width=0.78\columnwidth]{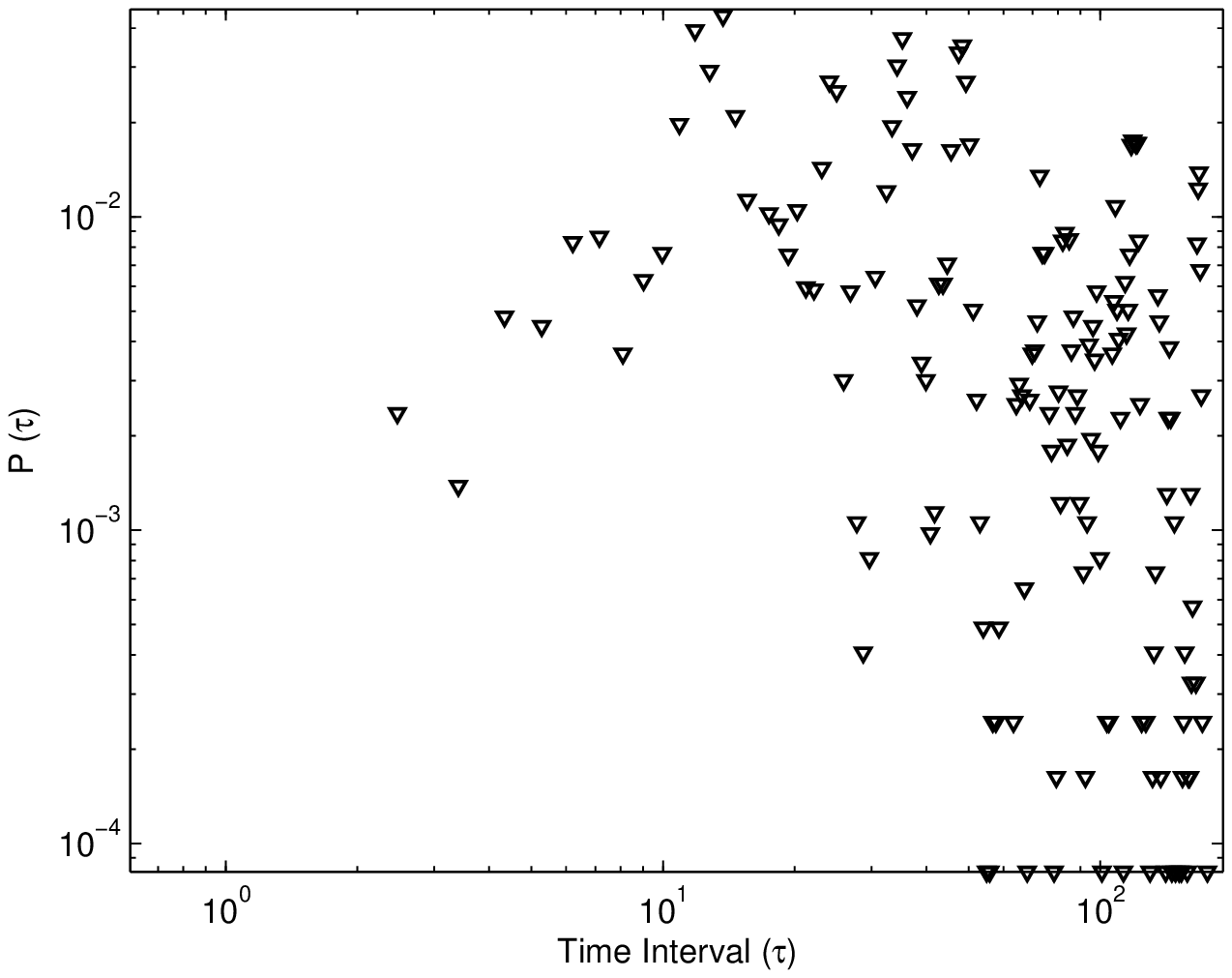}
\end{center}
\caption{SSR}
\label{9}
\end{figure}

%%%%%%%%%%%%%%%%%%%%%%%%%%%%%%%%%%%%%%%%%%%%%%%%%%%%%%%%%%%%
\begin{figure}[h!]
\begin{center}
\includegraphics[width=0.78\columnwidth]{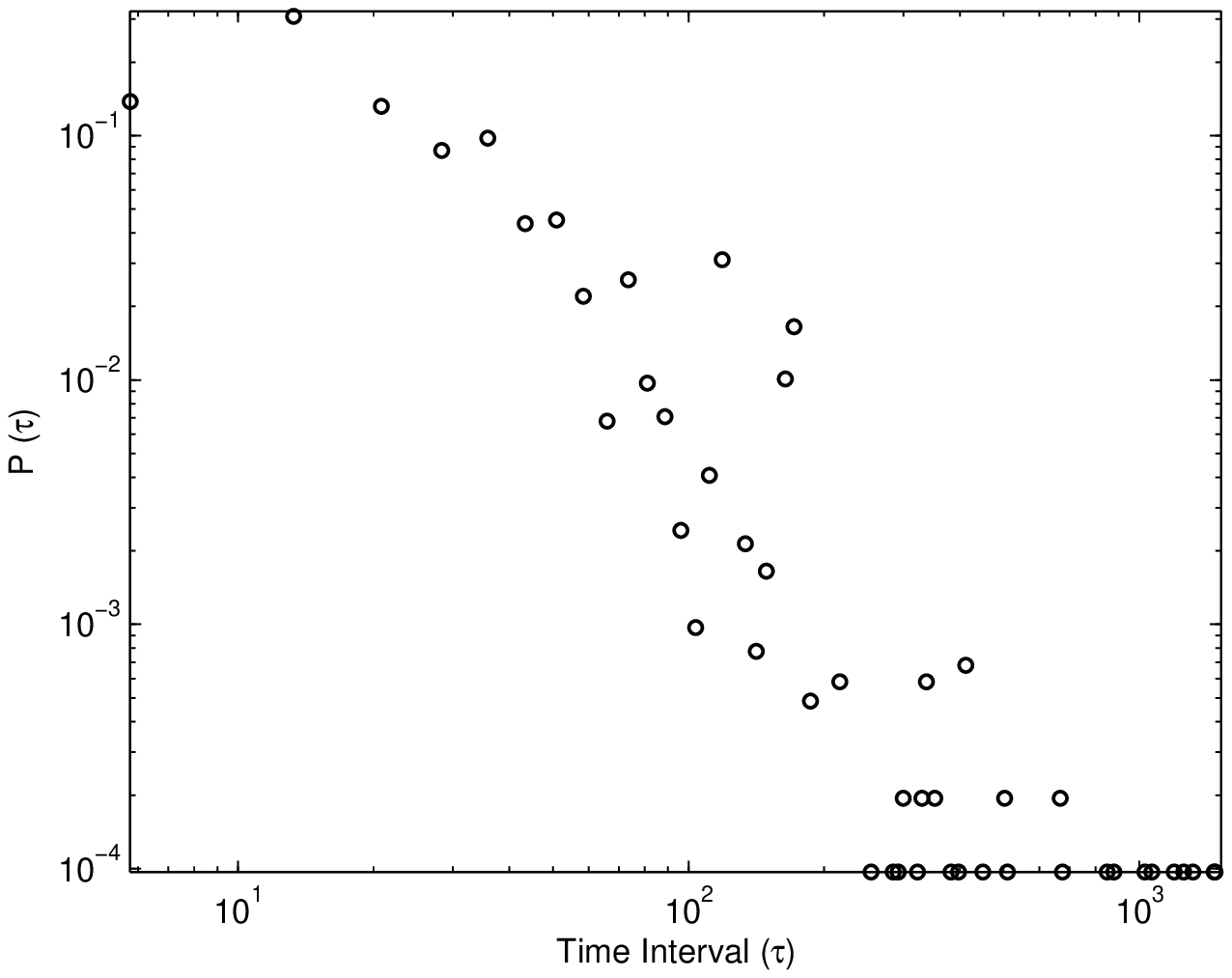}
\end{center}
\caption{SRS}
\label{10}
\end{figure}
%%%%%%%%%%%%%%%%%%%%%%%%%%%%%%%%%%%%%%%%%%%%%%%%%%%%%%%%%%%%%
%%%%%%%%%%%%%%%%%%%%%%%%%%%%%%%%%%%%%%%%%%%%%%%%%%%%%%%%%%%%
\begin{figure}[h!]
\begin{center}
\includegraphics[width=0.78\columnwidth]{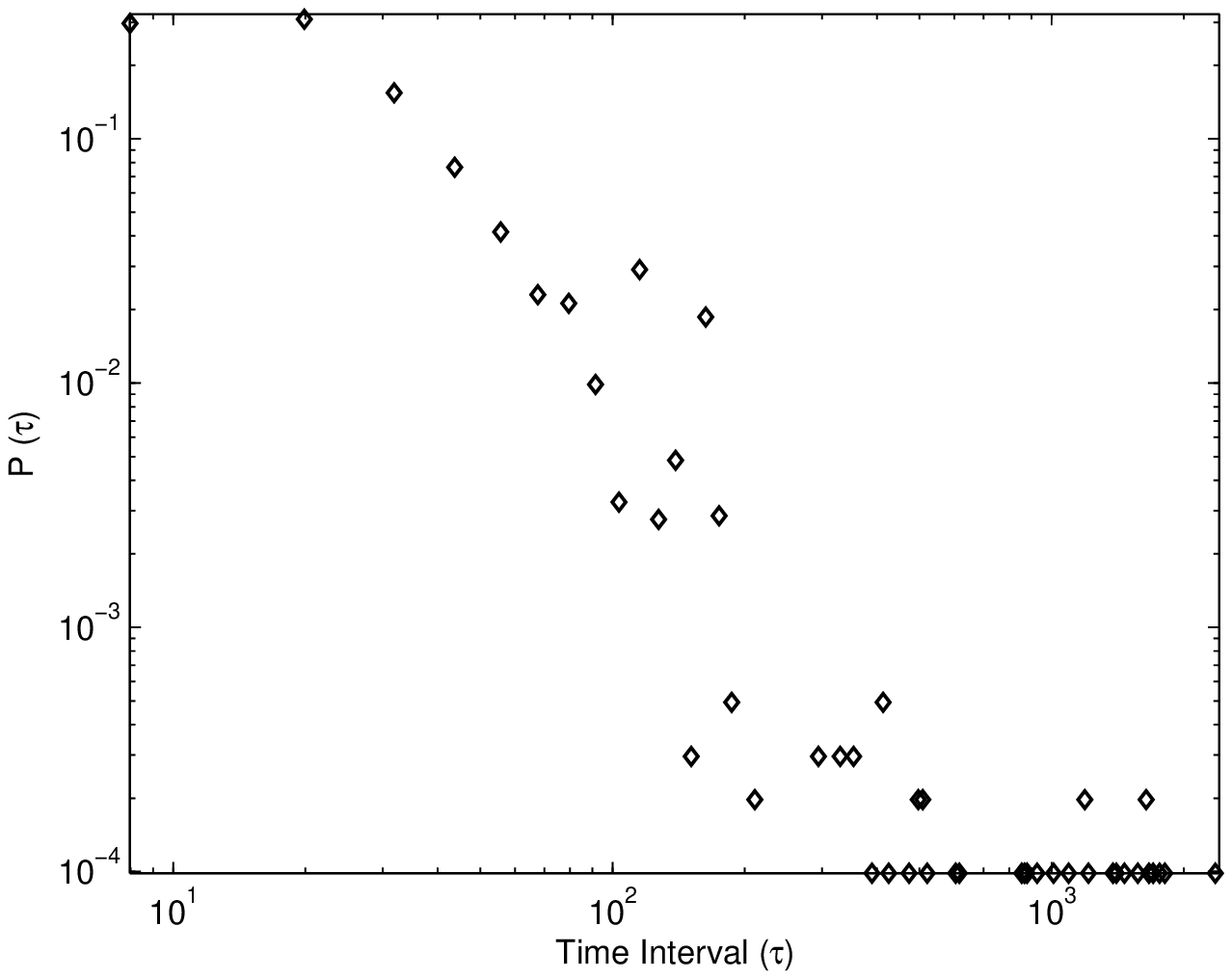}
\end{center}
\caption{SRR}
\label{11}
\end{figure}
%%%%%%%%%%%%%%%%%%%%%%%%%%%%%%%%%%%%%%%%%%%%%%%%%%%%%%%%%%%%%

In other words, music time series of notes played with constant pitch (no ``melody''), constant loudness (same intensities) but random rhythm (durations) do not exhibit the power law behavior. Music time series of notes played with constant pitch (no ``melody''), constant loudness (same intensities) and constant durations also do not exhibit this power law behavior. This study points towards the possibility that random characteristics of melody and loudness play an important role in the origin of this power law behavior of interval distributions in music while randomness of durations does not.

  %%%%%%%%%%%%%%%%%%%%%%%%%%%%%%%%%%%%%%%%%%%%%%%%%%%%%%%%%%%%%%%%%
\section*{\label{sec3} III. Conclusions}
%%%%%%%%%%%%%%%%%%%%%%%%%%%%%%%%%%%%%%%%%%%%%%%%%%%%%%%%%%%%%%%%%
In this paper, the recurrence interval between peaks in music time series is analyzed. Power law behavior P($\tau$) $\sim$ $\tau^{-\beta}$ is observed with the index $\beta$ varying from 1.5 to 2.7. The result is interesting in the light of the $1/f$ spectral density reported earlier by Voss and Clarke.  It has been shown earlier that a time series with uniform amplitudes and a $\tau^{-2}$  recurrence time distribution shows $1/f$ spectral density. Hence our results on interval distribution are in general agreement with the $1/f$ spectral density for music.

An attempt is made to study the origin of the above correlations in music. The role of melody, loudness variations, and rhythm in giving rise to these correlations is studied. The results seem to indicate that randomness characteristics of melody and loudness variations contribute more to the origin of the power law behavior than randomness of rhythm (or durations).

It is interesting to observe that the power law regime ends at about 20 milliseconds in all the cases studied. The origin of this cutoff may shed interesting light on the temporal correlations in music. The correlations studied in this paper are only between nearest peaks. One could study higher order correlations between more than two peaks.

%%%%%%%%%%%%%%%%%%%%%%%%%%%%%%%%%%%%%%%%%%%%%%%%%%%%%%%%%%%%

\begin{acknowledgments}
 We thank Ravishankar Sundararaman for writing the C program to generate the random music files and also for insightful discussions on the topic. We also thank Harshwardhan Wanare, Subhendu Ghosh, and V. Subrahmanyam for useful discussions at various stages of this work. Discussions about music, especially with regards to the nature of rhythm, with Deepak Gopinath were very useful and are appreciated.  
\end{acknowledgments}
%%%%%%%%%%%%%%%%%%%%%%%%%%%%%%%%%%%%%%%%%%%%%%%%%%%%%%%%%%%%%%%%%%%%

%%%%%%%%%%%%%%%%%%%%%%%%%%%%%%%%%%%%%%%%%%%%%%%%%%%%%%%%%%%%%%%%%%%%%%%%
%   References       
%%%%%%%%%%%%%%%%%%%%%%%%%%%%%%%%%%%%%%%%%%%%%%%%%%%%%%%%%%%%%%%%%%%%%%%%

\end{document}